\documentclass[
 aps,
 prc,
 preprint,
 showpacs,
superscriptaddress,
nofootinbib]{revtex4-1}
\usepackage{amsmath}
\usepackage{epsfig}
\usepackage{epstopdf}
\usepackage{graphicx}
\usepackage{soul}
\usepackage{color}
\usepackage{wasysym}
\usepackage[normalem]{ulem}
\usepackage{verbatim}
\usepackage{lipsum}
\usepackage{url}
\newcommand{\SO}{${}^1\textrm{S}_0\;$}
\newcommand{\SD}{${}^3\textrm{SD}_1\;$}
\newcommand{\PF}{${}^3\textrm{PF}_2\;$}

\begin{document}

\title{
Di--neutrons in neutron matter within Brueckner--Hartree--Fock approach
}

\author{Felipe Isaule}
\address{
  Department of Physics -- FCFM, University of Chile,
  Av. Blanco Encalada 2008, Santiago, Chile}
\author{H. F.  Arellano}
\address{
  Department of Physics -- FCFM, University of Chile,
  Av. Blanco Encalada 2008, Santiago, Chile}
\address{CEA,DAM,DIF F-91297 Arpajon, France}
\author{Arnau Rios}
\address{Department of Physics, 
             Faculty of Engineering and Physical Sciences, 
             University of Surrey, Guildford, 
             Surrey GU2 7XH, United Kingdom}

\date{\today}
\begin{abstract}
We investigate the appearance of di--neutron bound states in 
pure neutron matter within the Brueckner--Hartree--Fock approach 
at zero temperature. 
We consider the Argonne $v_{18}$ and Paris bare interactions 
as well as chiral two-- and three--nucleon forces.
Self--consistent single--particle potentials are calculated by
controlling explicitly singularities in the $g$ matrix associated 
with bound states.
Di--neutrons are loosely bound, with binding energies below $1$ MeV, 
but are unambiguously present for Fermi momenta below $1$~fm$^{-1}$ 
for all interactions. 
Within the same framework we are able to calculate and characterize 
di--neutron bound states, obtaining mean radii as high as $\sim\!110$~fm. 
Implications of these findings are presented and discussed.
\end{abstract}

\maketitle

\section{Introduction}
Di--neutrons have been investigated in a variety of regimes
and their imprint is recognized in the low--energy behavior
of phase shifts in neutron--neutron 
scattering \cite{Kobayashi2013,Spyrou2012,Kanada2008,Hammer2014208}. 
Bound and resonant states in few--neutron systems have been discussed both 
in recent experiments \cite{Kisamori2016} and in the 
theoretical nuclear physics literature \cite{Pieper2003}. 
Although the interaction between two neutrons in free space is attractive,
its strength is not enough to counterbalance the kinetic 
contribution of a confined system,
as implied by Heisenberg's uncertainty principle.
However, early in the 70s Migdal conjectured the possibility of
di--neutrons in the nuclear medium \cite{Migdal1972}, opening the way to
copious research on neutron pairing, superfluidity and clustering
in nuclear systems \cite{Schmidt1990,Typel2010,Broglia13,Ropke2015}.

The study we present in this work focuses on di--neutron
bound--state structures in homogeneous neutron matter
in the context of the Brueckner--Hartree--Fock (BHF) approach,
when realistic internucleon bare interactions are used.
Because of this, it offers the possibility
to be solved \textsl{ab initio} without additional parameters.
A novel element in this study is the accurate account 
for di--nucleon contributions in the evaluation of single--particle (sp)
spectra, as recently reported in Ref. \cite{Arellano2015} by one of us
for the case of symmetric nuclear matter based on the
Argonne $v_{18}$ internucleon bare interaction (AV18).
The three major conclusions of that work are:
\textsl{i.-} 
Nucleon effective masses can reach up to four times the bare nucleon mass;
\textsl{ii.-} 
Large size bound states occur at subsaturation densities; and
\textsl{iii.-} 
Two different families of sp solutions are found to 
satisfy self--consistency at low densities, leading to what is 
known as coexisting sp spectra.

The work we present here aims at re--assessing these properties in the
context of pure neutron matter, i.e. restricting the nucleon--nucleon
(\emph{NN}) interaction to unit isospin ($T=1$).
Microscopic studies of di--neutron formation in neutron matter, 
with many--body methods and based on realistic 
internucleon interactions, 
have not been discussed extensively in the past. 
Di--neutron states, 
if they appear in specific conditions within a neutron star's life,
could have an impact on its properties.
Hence, one would still like to assess to what extent the picture 
that arises when \emph{in-medium} di--nucleons are taken into account 
is robust under the use of realistic internucleon bare
interactions, i.e. independent of two--nucleon (2N) and
three--nucleon (3N) forces.
Furthermore, it would also be of interest to know what features
of di--nucleon structures occurring in
symmetric nuclear matter remain in the limit of pure neutron matter.

The study of neutron  matter properties over a 
wide range of densities is one of the quests of 
modern nuclear theory \cite{Gandolfi2015}.
At low densities, neutron matter may provide important clues 
for the understanding of neutron--rich nuclei, surface--related
nuclear phenomena and clustering \cite{Typel2010}.
At neutron densities near and above nuclear saturation, 
superfluid pairing is recognized as a key element to describe
the cooling evolution of neutron stars \cite{Broglia13}.
Additionally, the equation of state (EoS) for nuclear matter plays 
a central role in the description of hydrostatic equilibrium of
neutron stars \cite{Shapiro1983}.
With these considerations in mind, 
we supplement such studies with
ab initio calculations for \emph{in-medium} di--neutron bound states, 
which represent the first step towards a full microscopic
theory of clustering in dense matter. Our calculations only include
two--body clusters, but do so in a theoretically consistent way that does
not require phenomenological parameters \cite{Hempel2011}. 
In particular, our results could provide input for the more 
sophisticated cluster models that are employed in astrophysical 
simulations \cite{Typel2010,Hempel2011,Shen2011}.

The appearance of two--body bound states in dense, correlated media is 
key problem of interest to quantum many--body theory. 
The competition, or crossover,
between dimers and Cooper pairs has been discussed extensively in the
cold atom community \cite{Zwerger2011}. 
In neutron matter, a crossover of the same
type is also expected to 
appear \cite{Alm1993,Margueron2007,Jin2010,Stein2014}. 
While a detailed discussion of the crossover goes beyond the 
scope of our preliminary analysis, we point out some similarities 
between the di--neutron bound states and Cooper pairs, 
which point at a subtle relation between 
the two \cite{Clark2015,Stein2016,Khodel2014}.

This article is organized as follows.
In Sec. II, we lay out the theoretical framework under which di--nucleons
shall be investigated.
In Sec. III, we present results for sp spectra in the case
of neutron matter using realistic \emph{NN} potentials.
Furthermore, we examine the occurrence of di--neutrons, 
their spatial properties and effective masses.
We also provide a comparison of di--neutron bound states with
Cooper pairs and superfluid solutions in pure neutron matter.
As an application of our findings we also investigate the EoS of 
neutron matter.
Finally, we present a summary and conclusions of the work in Sec. IV .

\section{Framework}
The BHF approach for interacting nucleons in nuclear matter
can be identified as the lowest--order approximation of 
Brueckner--Bethe--Goldstone theory or self--consistent Green's 
function theory at
zero temperature \cite{Dickhoff2008}. 
In either case,
in the context of homogeneous nuclear matter,
the effective interaction is described
by a $G$ matrix, an operator that
depends on the density of the medium, and a starting energy $\omega$.
When only 2N correlations are taken into account
in the ladder approximation, the $G$ matrix satisfies
\begin{align}
\label{bhf}
G(\omega)=&v+v\, 
\sum_{\boldsymbol{k}_a,\boldsymbol{k}_b} 
|\boldsymbol{k_a k_b} \rangle \nonumber \times \\
 & \frac{[1-n(k_a)][(1-n(k_b)]}{\omega+i\eta-e(k_a) - e(k_b)} \;
\langle \boldsymbol{k_a k_b} |
\,G(\omega)\,,
\end{align}
where $v$ corresponds to the bare interaction between nucleons;
$e(k)\equiv e_k$, 
is the sp energy given in terms of the sp potential, $U(k)$,
\begin{equation}
\label{esp}
e_k=\frac{k^2}{2m} + U(k)\,;
\end{equation}
and $n(k)$ corresponds to the momentum distribution.
The density is obtained from the latter through the relation
\begin{equation}
  \label{nk}
  \rho = \nu\sum_{\boldsymbol{k}} n(k)\,,
\end{equation}
where $\nu=2(4)$, for neutron (nuclear) matter.
At zero temperature, the momentum distribution is a step function,
 $n(k)=\Theta(k_F-k)$, with $k_F$ being the Fermi momentum,
\begin{equation}
  \label{kf}
  \rho = \nu\,\frac{k_F^3}{6\pi^2}\;.
\end{equation}

The $G$--matrix in Eq.~(\ref{bhf}) enables 
the evaluation of the sp potential,
\begin{equation}
\label{usp}
U(k)=\textsf{Re}\, \sum_{\boldsymbol{p}} n(p) 
\langle
\textstyle{ \frac{\boldsymbol{k}-\boldsymbol{p}}{2}}
|g_{|\boldsymbol{k+p}|}(e_k + e_p) | 
\textstyle{ \frac{\boldsymbol{k}-\boldsymbol{p}}{2}}
\rangle_A\;,
\end{equation}
where $g_K(\omega)$ represents the reduced $G(\omega)$ matrix 
for total pair momentum $\vec{K}=\vec{k}+\vec{p}$, 
after a momentum--conserving Dirac $\delta$ function has been factored out.
The subscript \emph{A} denotes the antisymmetrization of the ket state.
Self--consistency of Eqs. (\ref{bhf}), (\ref{esp}) and (\ref{usp}) 
is achieved iteratively. 
We use of the continuous choice for the sp spectrum, 
where the condition expressed by Eq. (\ref{usp}) is imposed 
on all momenta $k$. This choice of sp potential yields better 
convergence in terms of the hole--line expansion \cite{Song98}.

In the iterative procedure to determine $U(k)$, actual bound states 
appear while evaluating Fermi--motion integrals in Eq. (\ref{usp}).
As found in symmetric nuclear matter,
the appearance of bound states is more relevant at subsaturation 
densities~\cite{Arellano2015}, where di--nucleon singularities take 
place in both the \SO and \SD channels.
This feature has resulted in the identification of
two distinct families of sp solutions meeting self--consistency:
Phase I, for $0\leq k_F\lesssim 0.29$~fm$^{-1}$; and
Phase II, for $k_F\gtrsim 0.13$~fm$^{-1}$.
Over the range $0.13\lesssim k_F\lesssim 0.29$~fm$^{-1}$, the
two families satisfy self--consistency criteria, and are referred to 
as ``coexisting solutions'' in Ref. \cite{Arellano2015}.
The scenario for pure neutron matter is different, since the isoscalar 
channel is suppressed, 
ruling out contributions to the sp potential from the \SD state.
As a result, we find that pure neutron matter does not allow for
coexisting sp spectra.

An appealing feature of the BHF approach is that 
it enables a direct link between the bare interaction 
among constituents and properties of interacting nucleons 
in a homogeneous environment. 
Microscopic, phase--shift equivalent nuclear interactions predict
different \emph{in-medium} properties \cite{Baldo2012}. In this work, 
we explore a variety of nuclear interactions in an effort to 
provide robust predictions for di--neutron properties. We use the
traditional 
Argonne $v_{18}$ \cite{Wiringa1995} and Paris \cite{Paris} bare
potentials, which have been fit to \textsl{NN} phase--shift data
at beam energies below pion production threshold, 
together with static properties of the deuteron.
Additionally, we also include a chiral 
effective--field--theory ($\chi$EFT) interaction, which is based on chiral 
perturbation theory.
The resulting bare interaction is constructed with nucleons and
pions as degrees of freedom, with the 2N part fit
to two--nucleon data. We consider the chiral 2N force (2NF) up to 
next--to--next--to--next--to--leading order (N$^3$LO) given
by Entem and Machleidt \cite{Entem2003}.
In addition, we also consider chiral 3\textsl{N} forces (3NF) in
N$^2$LO, using a density--dependent 2NF at the two--body level
\cite{Holt2010,Hebeler2010a}. This density--dependent contribution
does not contain correlation effects \cite{Carbone2014}, 
and is added to the bare 2NF 
in the calculation of the $G$--matrix. The corresponding Hartree--Fock 
contribution is subtracted at each iteration to avoid any double counting. 
For this chiral 3NF contribution,
we use the low energy constants $c_D = -1.11$ and $c_E = -0.66$,
reported in Ref. \cite{Nogga2006}, which describe the $^3$H and
$^4$He binding energies with unevolved \emph{NN} interactions. 
We also note that we have used a chiral cutoff of $\Lambda_\chi = 700$~MeV.
Because the main aim of our work is not the estimation of systematic effects, 
we do not explore the cutoff or low--energy--constant dependence 
of our results. Instead,
we will focus most of our analysis on the region where all predictions for 
di--neutron bound states are robust 
(or invariant with respect to the underlying potential).
Roughly speaking, we find that results with different interactions 
are equivalent 
below $k_F=1.5$~fm$^{-1}$.

\section{Results}
We have obtained self--consistent solutions for sp potentials 
for pure neutron matter in the BHF approximation within the 
continuous choice following the techniques 
described in Ref. \cite{Arellano2015}. 
Calculations were made by considering all partial waves up to $J=7$ 
in the total angular momentum, regardless of the density. 
Di--nucleon occurrences in the \SO and \PF \emph{NN} channels 
are explicitly treated.
Files containing self--consistent sp potentials discussed in 
this work can be retrieved from Ref.~\cite{omponline}.

\subsection{Solutions}

\begin{figure}[ht!]
 \centering
 \includegraphics[width=0.8\linewidth]{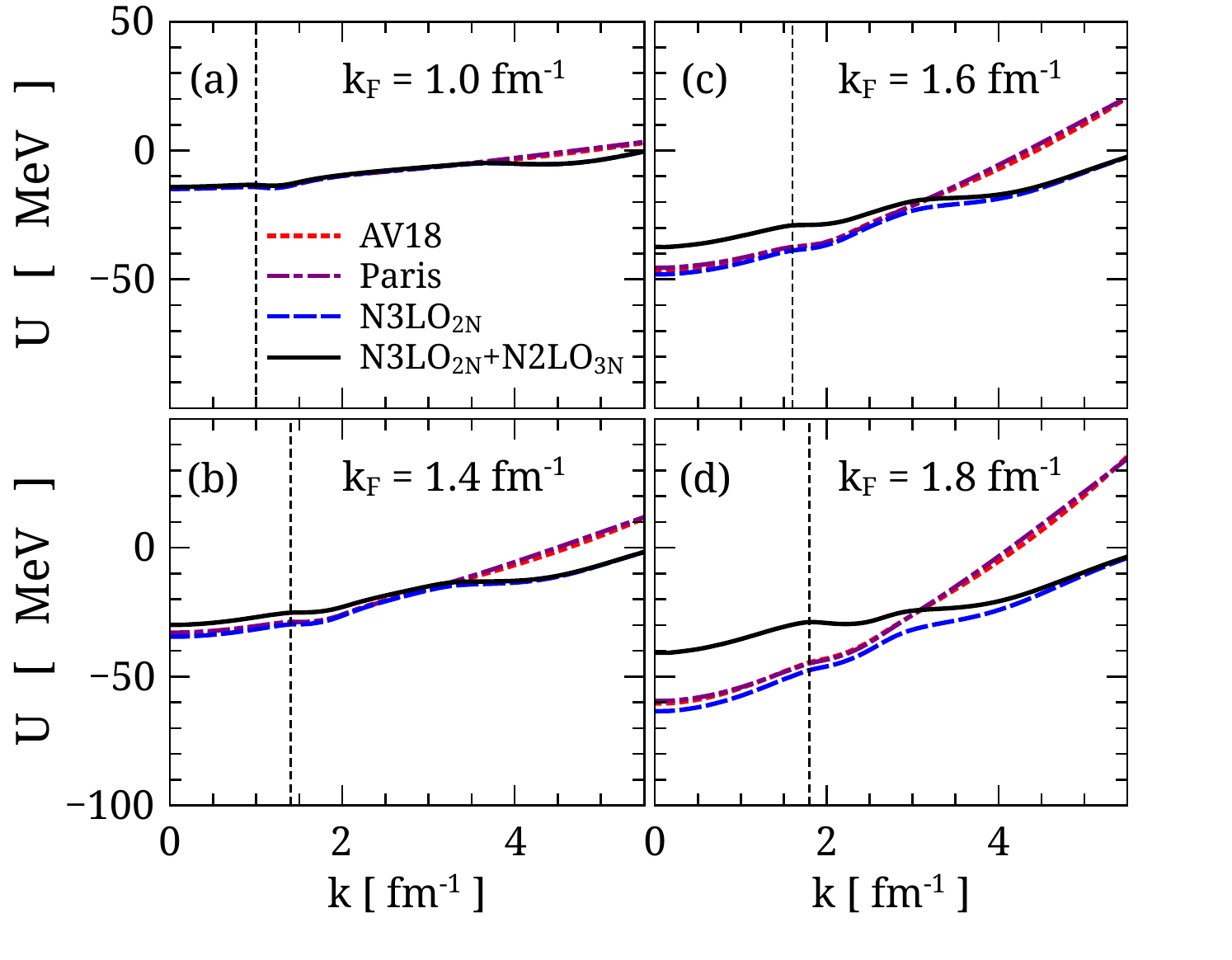}
 \caption{\label{fig_bhf}
   BHF self--consistent sp potentials for pure neutron matter as a
   functions of sp momentum, $k$.  
   Short--dashed, dash--dotted, dashed and solid curves correspond to 
   AV18, Paris, N3LO$_{2N}$ and N3LO$_{2N}$+N2LO$_{3N}$ 
   bare potentials, respectively. 
   Panels (a)-(d) present solutions for 
   $k_F=1.0,\;1.4,\; 1.6,$ and 1.8~fm$^{-1}$, respectively.}
\end{figure}

In Fig. \ref{fig_bhf}, we show self--consistent solutions $U(k)$ for 
pure neutron matter as a function of sp momentum, $k$. 
We include results corresponding to AV18 (short--dashed curves), 
Paris (dash--dotted curves),
N3LO$_{2N}$ (dashed curves) and N3LO$_{2N}$+N2LO$_{3N}$ (solid curves). 
Figures 1(a)-1(d) represent solutions for 
$k_F=1.0,\;1.4,\; 1.6,$ and 1.8~fm$^{-1}$, respectively. 
These Fermi momenta correspond to neutron densities between
0.03 and 0.2~fm$^{-3}$, with the latter close to the saturation
density of symmetric nuclear matter.
Dashed vertical lines in each panel represent the corresponding 
Fermi momentum.
We keep the same vertical scale throughout for clarity.
The results shown in Fig.~1(a) for $k_F=1$~fm$^{-1}$ provide evidence for
nearly identical behavior for all four interactions 
for  $k<3.5$~fm$^{-1}$.
This similarity points to a dominance of many--body correlations in the 
low density regime, so that individual \emph{NN} force features are not 
explicitly resolved. 

In Figs.~1(c) and 1(d), where $k_F\geq 1.6$~fm$^{-1}$, 
we identify two aspects of interest. On the one hand, 
for momenta below about $k\lesssim 2$~fm$^{-1}$, 
the sp solutions based on N3LO$_{2N}$ 
follow the trend given by AV18 and Paris interactions. 
The ``low--momentum'' part of $U(k)$ becomes more
and more attractive with density, and is mildly dependent
on momentum up to the corresponding $k_F$. 
As long as 2NF are used, this low--momentum
components are relatively independent of the nuclear interaction. 
In contrast,
the sp solution based on the N3LO$_{2N}$+N2LO$_{3N}$ interaction
is substantially more repulsive at low momenta. This agrees with the
intuitive idea that 3NF provide repulsion in the $T=1$ channel 
\cite{Hebeler2010a}. We note in particular that the low--momentum 
part of $U(k)$ becomes more and more repulsive as the density increases 
when 3NF are considered.

On the other hand, for $k\gtrsim 3$~fm$^{-1}$ the sp solutions 
for the two chiral interactions get closer to each other 
as momentum increases, 
departing markedly from the trend based on AV18 and Paris potentials.  
Whereas traditional interactions have
strong short--range cores, chiral forces are cut off by construction. 
As a consequence, there is no support in the bare interaction for any 
high--momentum components of $U(k)$. 
In particular, 
one can see that a sharp relative momentum cutoff at $\Lambda$ 
will affect sp spectra at sp momenta $k>2\Lambda-k_F$. 
The Entem--Machleidt 
interaction has $\Lambda=500$~MeV ($\sim\!2.5$~fm$^{-1}$), 
and hence the similarity
between the two forces beyond about $3$~fm$^{-1}$ can be ascribed 
to the disappearance of the matrix elements at high momenta. 
In contrast, ``hard core'' forces like Paris and AV18 still 
have active components at large $k$, giving rise to nonzero 
spectra at arbitrarily large momenta.
   
\begin{figure}[ht!]
  \centering
    \includegraphics[width=0.8\linewidth]{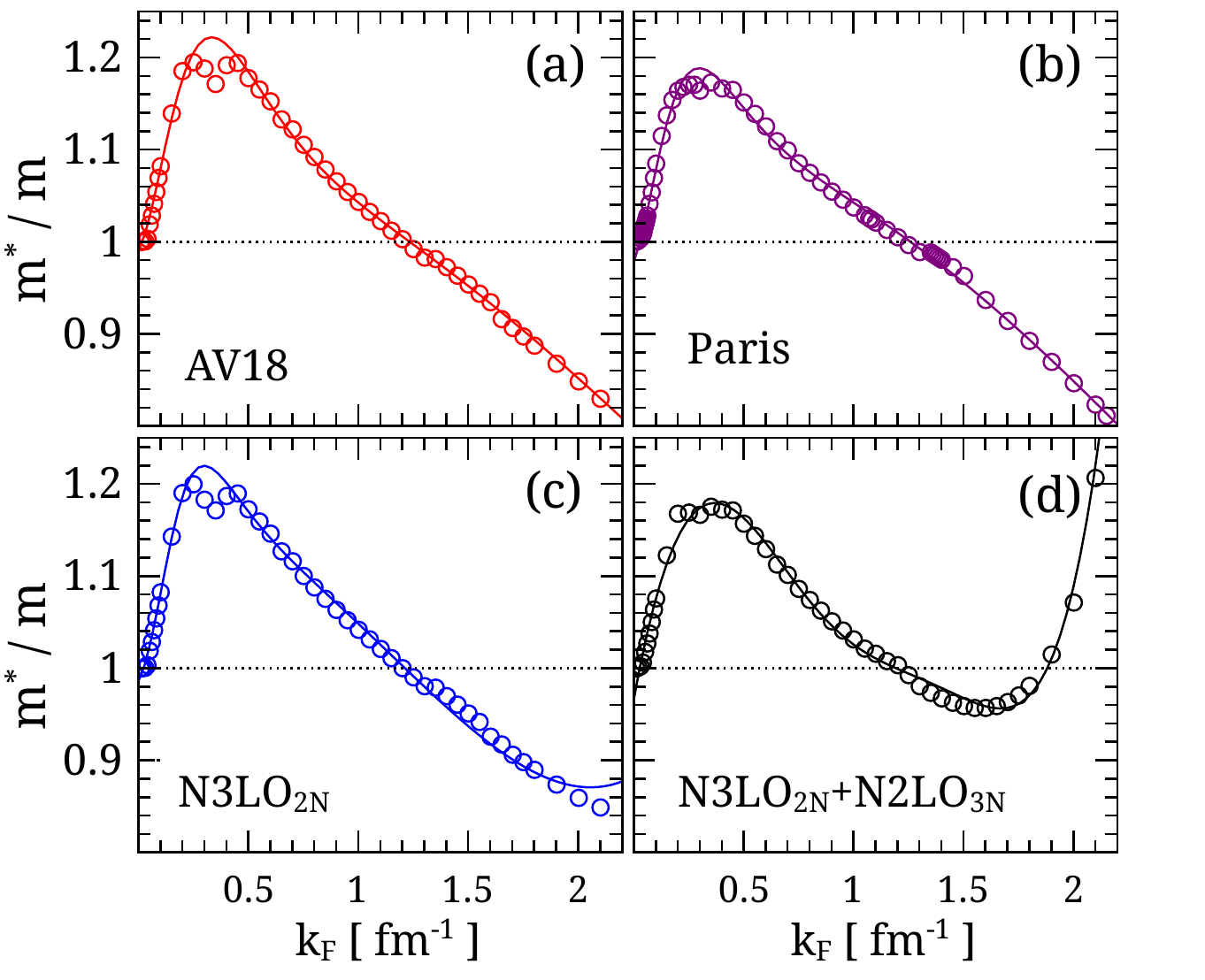}
   \caption{\label{effectivemass}
Effective--to--bare--mass ratio $m^*/m$ in pure neutron matter 
as a function of Fermi momentum, $k_F$.
Panels (a)-(d) correspond to results for 
AV18, Paris, N3LO$_{2N}$, and
N3LO$_{2N}$+N2LO$_{3N}$ bare potentials, respectively.
Continuous curves represent fits based on Eq.~(\ref{MassPar}).
}
\end{figure}

The study of neutron effective masses $m^{*}$ in both 
low-- and high--density neutron matter has been a subject of 
increasing interest over the past few years \cite{Chamel2013,Baldo2014}.
The calculated sp spectra of Eq.~(\ref{esp}) provide 
access to the effective mass,
\begin{equation}
\label{masseff}
\frac{m^*}{m} = 
\frac{k_F}{m}\left [
\frac{\partial e_k}{\partial k} \right ]^{-1}_{k=k_F}\,,
\end{equation}
where $m$ stands for nucleon bare mass.
We show in Fig.~\ref{effectivemass} the calculated 
effective--to--bare mass ratio $m^*/m$ as a function of 
Fermi momentum for the four interactions discussed above.   
Figures 2(a)-2(d) correspond to results based
on AV18, Paris, N3LO, and N3LO$_{2N}$+N2LO$_{3N}$, respectively.
We focus our discussion around two basic issues. 
First, the ratio $m^*/m$ for the four interactions follows 
almost identical behaviors for Fermi momenta below $1.5$~fm$^{-1}$. 
We take this as an indication of the robustness of our results. 
Starting, as expected, from a value of $m^*=m$ at low density, 
the effective mass grows
with $k_F$ and reaches a maximum of around $\sim\!1.2 m$ 
at Fermi momenta in the range $0.3-0.4$~fm$^{-1}$. From this point
on, the effective mass decreases and becomes smaller than $m$ 
above $k_F \approx 1$~fm$^{-1}$. As we shall discuss in the following, 
the density regime where $m^*>m$ is nearly the same 
as that where di--neutron bound states take place. 
Furthermore, in the context of the sp potentials of Fig.~\ref{fig_bhf}, 
we emphasize that $m^*>m$ states will be found whenever the sp potential 
is a decreasing function of momentum around $k_F$. This is indeed the case
in Figs. 1(a) and 1.(b),
where one can identify a wiggly structure 
around $k_F$. 
We note that effective masses at such low densities are relevant for the 
physics of neutron star crusts, which plays an important role
in the description of pulsar glitches by entrainment 
effects \cite{Chamel2006263,Chamel2013}. 

Second, we find that effective masses based on 2NF--only calculations
tend to decrease with $k_F$ above the Fermi momentum of $1.5$~fm$^{-1}$.
In contrast, the ratio $m^*/m$ for N3LO$_{2N}$+N2LO$_{3N}$ starts to 
grow in this region, while the other three cases decrease monotonically. 
As a matter of fact, we find a region where 
$m^{*}/m>1$  for $k_F$ above $1.8$~fm$^{-1}$. 
Differences in the effective masses between chiral 
2NF and 2NF+3NF calculations 
for $k_F>1.5$~fm$^{-1}$ have already been reported 
in Ref. \cite{Hebeler2010a}. As mentioned earlier, values of $m^*/m$ above
1 indicate the presence of decreasing sp potentials as a function of $k$. 
Figure 1(d) illustrates the appearance of such a 
structure for the N3LO$_{2N}$+N2LO$_{3N}$ calculations. 
We attribute this decrease to the competition between the 
strong repulsion induced by 3NF at low momenta and the onset of cutoff 
effects at momenta close to $\sim\!3$~fm$^{-1}$.
A detailed analysis of this behavior goes beyond the scope of this work,
because it would require a study of the low--energy--constant and cutoff 
dependence of these results. 

We found a suitable parametrization 
for the calculated effective masses in the form
\begin{equation}
\label{MassPar}
\frac{m*}{m} = 1+(a_0+a_2x^2)P(x)+(b_0+b_2x^2+b_4x^4)Q(x),
\end{equation}
where $x=k_F/b$, with $b = 0.2$~fm$^{-1}$ representing a typical 
Fermi momentum for the effective mass maximum. 
The functions $P$ and $Q$ satisfy $P+Q=1$, with
\begin{equation}
  P(x) = \frac12 \left [ 1 - \tanh\left(\frac{x-x_c}{d}\right )\right ]\;,
\end{equation}
which provides an effective separation for two regions 
of density dependence.
The resulting coefficients for each interaction 
are summarized in Table \ref{fitmasses}.  
The continuous curves in Fig.~\ref{effectivemass} show 
the resulting $m^*/m$ parametrizations.

\begin{table}[t!]
\centering
\begin{tabular}{  l|r r r r  }
\hline
&  AV18\hspace{18pt}
& Paris\hspace{20pt}
& N3LO$_{2N}$\hspace{16pt}
& N3LO$_{2N}$\hspace{16pt}  \\
&	&	&	& +N2LO$_{3N}$\hspace{16pt} \\
\hline
\hline
$x_c$ &
$0$\hspace{32pt}   & 
$0$\hspace{32pt}   & 
$0.6197$\hspace{32pt}  &
$5.390$\hspace{32pt}   \\
$d$ &
$1.411$\hspace{32pt}   &
$1.186$\hspace{32pt}   &
$0.9544$\hspace{32pt}  &
$4.193$\hspace{32pt}   \\
\hline
$a_0$ \rule{0pt}{3ex}  &
$ -1.158\times 10^{-1} $ &
$ -1.549\times 10^{-1} $ &
$ -6.633\times 10^{-2} $ &
$ -6.313\times 10^{-1} $ \\
$a_2$ &
$  6.299\times 10^{-1} $ &
$  6.245\times 10^{-1} $ &
$  2.905\times 10^{-1} $ &
$ -1.542\times 10^{-1} $ \\
$b_0$ &
$  9.668\times 10^{-2} $ &
$  1.104\times 10^{-1} $ &
$  1.744\times 10^{-1} $ &
$  7.771\times 10^{-1} $ \\
$b_2$ &
$ -2.762\times 10^{-3} $ &
$ -2.927\times 10^{-3} $ &
$ -5.751\times 10^{-3} $ &
$ -1.036\times 10^{-1} $ \\
$b_4$ \rule[-1.2ex]{0pt}{0pt} &
$  3.128\times 10^{-6} $ &
$  3.111\times 10^{-6} $ &
$  2.722\times 10^{-5} $ &
$  4.467\times 10^{-4} $ \\
\hline
\end{tabular}
\caption{Fit coefficients for the effective mass, $m^*/m$, 
  following Eq. (\ref{MassPar}). Fits were carried out in the range
$0.05\leq k_F\leq 3$~fm$^{-1}$ for all forces, 
except for N3LO$_{2N}$+N2LO$_{3N}$, 
where $0.05\leq k_F\leq 2.2$~fm$^{-1}$.
}
\label{fitmasses}
\end{table}

\subsection{Di--neutrons in pure neutron matter}

Di--neutrons constitute bound states of two neutrons. 
As such, they have an associated wave function of finite range
and eigenenergy below the Fermi surface.
Such states do not occur in free space, despite the attractive nature
of the interaction as indicated by the \SO scattering length. 
In the context of BHF equations,
the starting energy $\omega$ at which these states occur satisfies
the criterion
\begin{equation}
\label{det}
\det[1 - v \Lambda_K(\omega)]=0\;,
\end{equation}
with $\Lambda_K(\omega)$ the particle--particle propagator in 
Eq. (\ref{bhf}) for pairs with total momentum $K$. 
States are only bound below a corresponding \emph{in-medium} threshold, 
which depends on the sp energies at each iteration. In particular, 
bound states require that $\omega<\omega_{th}$, 
with $\omega_{th}$ being the threshold (lowest) sp energy of the pair 
allowed by Pauli blocking.  
In the particular case of pairs with their center of mass (c.m.) 
at rest ($K=0$), the energy threshold occurs at $\omega_{th}=2e(k_F)$. 
We search automatically for di--neutrons 
during the iterative BHF solving process. 

\begin{figure}[ht!]
  \centering
    \includegraphics[width=0.8\linewidth]{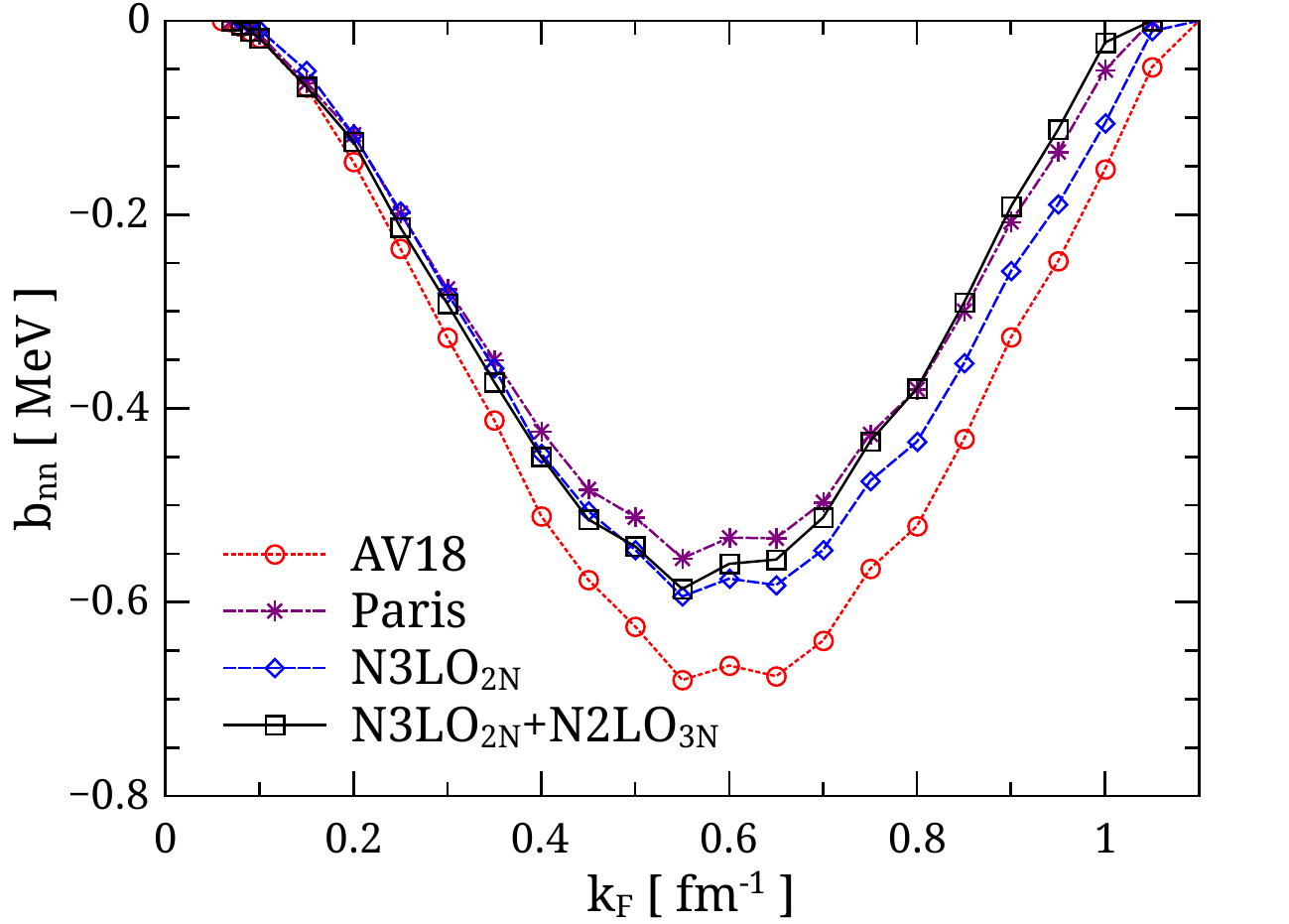}
   \caption{\label{fig_bnn}
Di--neutron binding energy in pure neutron matter as a 
function of Fermi momentum. 
Circles, asterisks, diamonds and squares 
correspond to AV18, Paris, N3LO$_{2N}$ and 
N3LO$_{2N}$+N2LO$_{3N}$ bare potentials, respectively.
Results correspond to pairs with zero total momentum.}
\end{figure}

Solutions of Eq. (\ref{det}) lead to a di--neutron binding 
energy $b_{nn}$ defined with respect to the corresponding threshold,
\begin{equation}
\label{bnn}
b_{nn} = \omega - \omega_{th}\;.
\end{equation}
Under this convention of sign, 
also adopted in Refs.~\cite{Witala2012,Sedrakian2006,Ropke1998},
a bound state has negative energy. 
In Fig. \ref{fig_bnn} we show
the di--neutron binding energy as a function of Fermi momentum.
These results correspond to static bound pairs, where $K=0$. 
Circles, asterisks, diamonds, and squares represent  AV18, Paris,
N3LO$_{2N}$ and N3LO$_{2N}$+N2LO$_{3N}$ interactions, respectively. 
One of the major conclusions of our work is that all four 
interactions considered hold bound di--neutrons states at 
Fermi momenta in the range $0.05-1.05$~fm$^{-1}$. 
Di--neutrons are relatively loosely bound, 
with binding energies below $700$ keV. 
In all cases, the deepest binding takes place 
at $k_F\approx 0.6$~fm$^{-1}$, 
with binding energies ranging between $550$ and $700$ keV.
AV18 yields the largest binding, $b_{nn}=-0.7$~MeV, 
whereas Paris and the chiral interactions produce slightly 
less bound di--neutrons, $b_{nn}\approx\, -0.6$~MeV.
Let us emphasize the fact that the prediction of the 
appearance of di--neutron bound states is extremely robust, 
in the sense that it is to a large extent independent of the 
interaction under consideration. 
We note, in particular, that both the binding energy and the density 
dependence of the di--neutron states are predicted to be very similar. 
This is a strong indication that the appearance of di--neutrons is 
governed by many--body correlations.

\begin{figure}[ht!]
  \centering
  \includegraphics[width=0.8\linewidth]{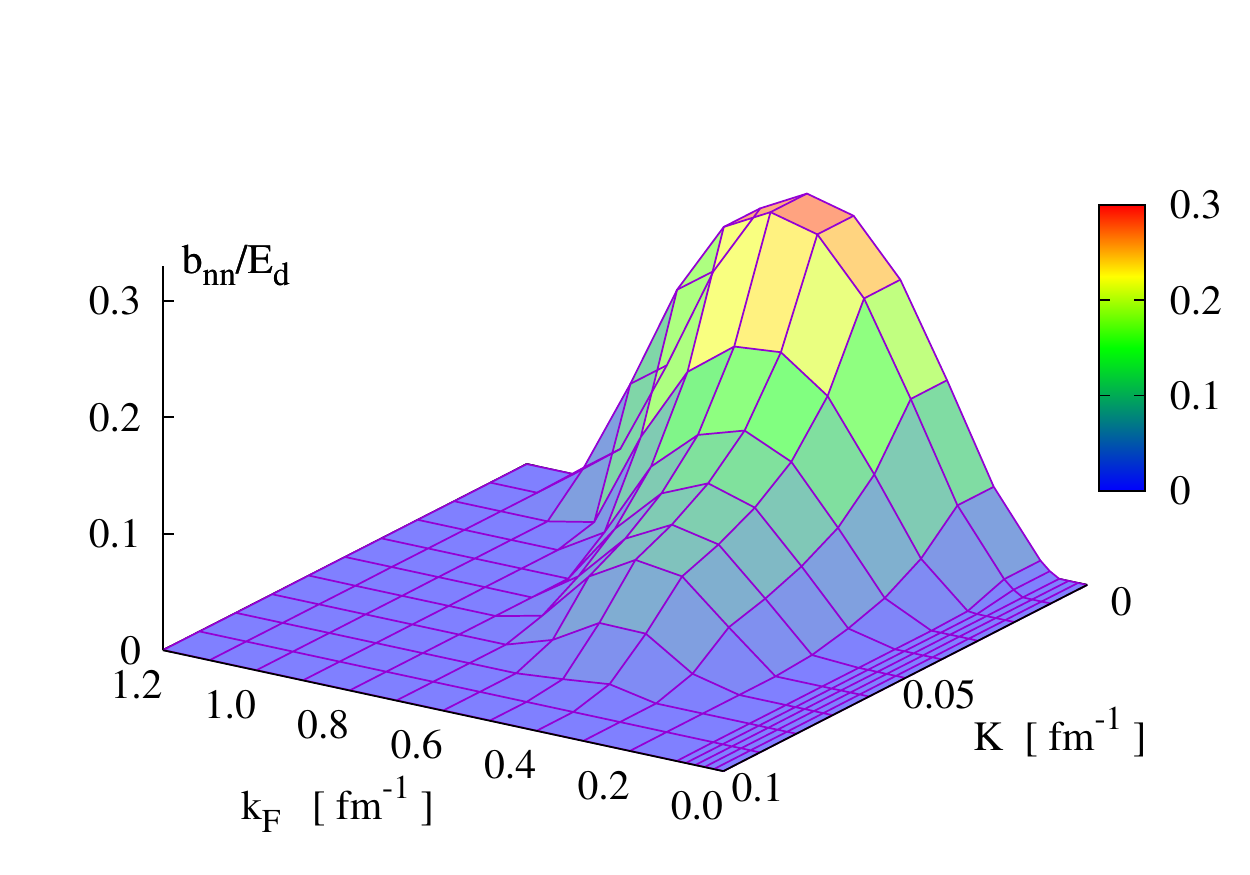}
  \caption{\label{fig_bnn3d}
Di--neutron binding energy, relative to the deuteron binding energy 
in vacuum, as a function of the pair momentum $K$ and $k_F$ 
in pure neutron matter.  Results have been obtained with the 
N3LO$_{2N}$+N2LO$_{3N}$ interactions.}
\end{figure}

The BHF approach offers the appealing possibility of 
obtaining di--nucleon bound states even beyond the static case. 
When $K \neq 0$, the c.m. of the pair 
moves relative to the nuclear medium, 
and thus one expects the binding to decrease as $K$ increases. 
We provide an illustrative example of this behavior 
in Fig.  \ref{fig_bnn3d}, where we show a surface plot for 
$b_{nn}/E_d$\footnote{Here, $E_d$ corresponds to the binding 
energy of the deuteron  in free space. }
as a function of $k_F$ and $K$.
The bare interaction considered in this case is 
N3LO$_{2N}$+N2LO$_{3N}$, 
but a similar behavior is found for the other three interactions.
In this case the strongest binding ($\sim\!630$~keV) 
occurs for static pairs ($K=0$) at $k_F\approx 0.6$~fm$^{-1}$. 
As the c.m. momentum increases, binding diminishes smoothly. 
At $k_F\approx 0.4-0.5$~fm$^{-1}$ di--neutrons take place over a larger
range of total momenta, allowing bound states up to
$K$ slightly below $\sim\!0.1$~fm$^{-1}$.
This bodes well with the idea that di--neutrons can propagate 
more easily in a relatively dilute and uncorrelated medium, 
before they are affected by Pauli blocking at large densities.
Note that for $k_F\to 0$, di--neutrons get dissolved, as should
be expected from realistic interactions that do not hold 
bound states in free space in the channel \SO.
This feature contrasts with the deuteron channel in symmetric nuclear 
matter, where a binding energy of -2.22~MeV is allowed for
all $K\geq 0$ in the case $k_F=0$.
On the other hand, as $k_F$ increasess, di--neutrons get dissolved.
As we shall discuss in the following sub--section, 
this feature appears related to the fact that the effective 
mass gets close to or smaller than the neutron bare mass.

\subsection{Bound state wave functions}
\label{dinucleons}

As we have just discussed, singularities of $g_K(\omega)$ in the 
\SO and \PF channels
on the real energy axis at energies below the threshold
(i.e. $\omega<\omega_{th}$) point to di--neutron bound states. 
In the following, we discuss the procedures that we have followed to 
obtain the associated \emph{in-medium} bound--state wave functions.
Let $\omega_{nn}$ be the eigenenergy corresponding to one such
bound state. 
One can show that, for starting energies close to the real axis, 
the $G$--matrix fulfills the equation \cite{Arellano2015}
\begin{equation}
\label{ieta}
\lim_{\eta\to 0} \,i\eta\, g_K(\omega_{nn}+i\eta)= 
vQ|\psi\rangle\langle\psi|Qv\;,
\end{equation}
with $Q$ being the Pauli blocking operator and $|\psi\rangle$ being the 
corresponding eigenstate.
On the other hand, consider the wave equation for $|\psi\rangle$
associated with Eq.~(\ref{bhf}),
\[
(\omega_{nn} - \hat e_1 -\hat e_2)| \psi \rangle = QvQ |\psi \rangle\;,
\]
from which we obtain the projected wave function in momentum space,
\begin{equation}
\label{psi}
\psi(q)=
\frac{\langle{q}|QvQ|\psi\rangle}{\omega_{nn}-E_K(q)}\;.
\end{equation}
Here, $E_K(q) = \langle e(k_a)+e(k_b)\rangle$, corresponds to the 
angle--averaged particle--particle energy when 
$\boldsymbol{k}_{a,b}=
\boldsymbol{K}\pm\textstyle{\frac12}{\boldsymbol{q}}$.
In the case of pairs without c.m. motion relative 
to the nuclear medium, the numerator reduces to
$\Theta(q-k_F) \langle{q}|vQ|\psi\rangle$,
indicating that di--nucleon wave functions in the nuclear medium
cannot contain momentum components that are already occupied 
by single particles.

The wave function $\Psi(q)$ can be obtained after 
$\langle{q}|vQ|\psi\rangle$ is extracted from Eq.~(\ref{ieta}). 
We achieve this numerically after solving $g_K(\omega_{nn}+i\eta)$
for a series of small values of $\eta$ and 
extrapolating $\eta\to 0$ afterwards.
In the case of uncoupled channels with orbital angular momentum $L$,
the wave function in coordinate space is obtained from the usual 
Bessel transform
\begin{equation}
\label{fourierL}
\Psi(r) = \sqrt{\frac{2}{\pi}} 
\int_{\bar q}^\infty q^2\,dq\,j_L(qr)\psi(q)\;.
\end{equation}
We note, however, that there is a lower integration bound, 
$\bar q$ that corresponds to that allowed
by Pauli blocking.

\begin{figure}[ht!]
  \centering
  \includegraphics[width=0.8\linewidth]{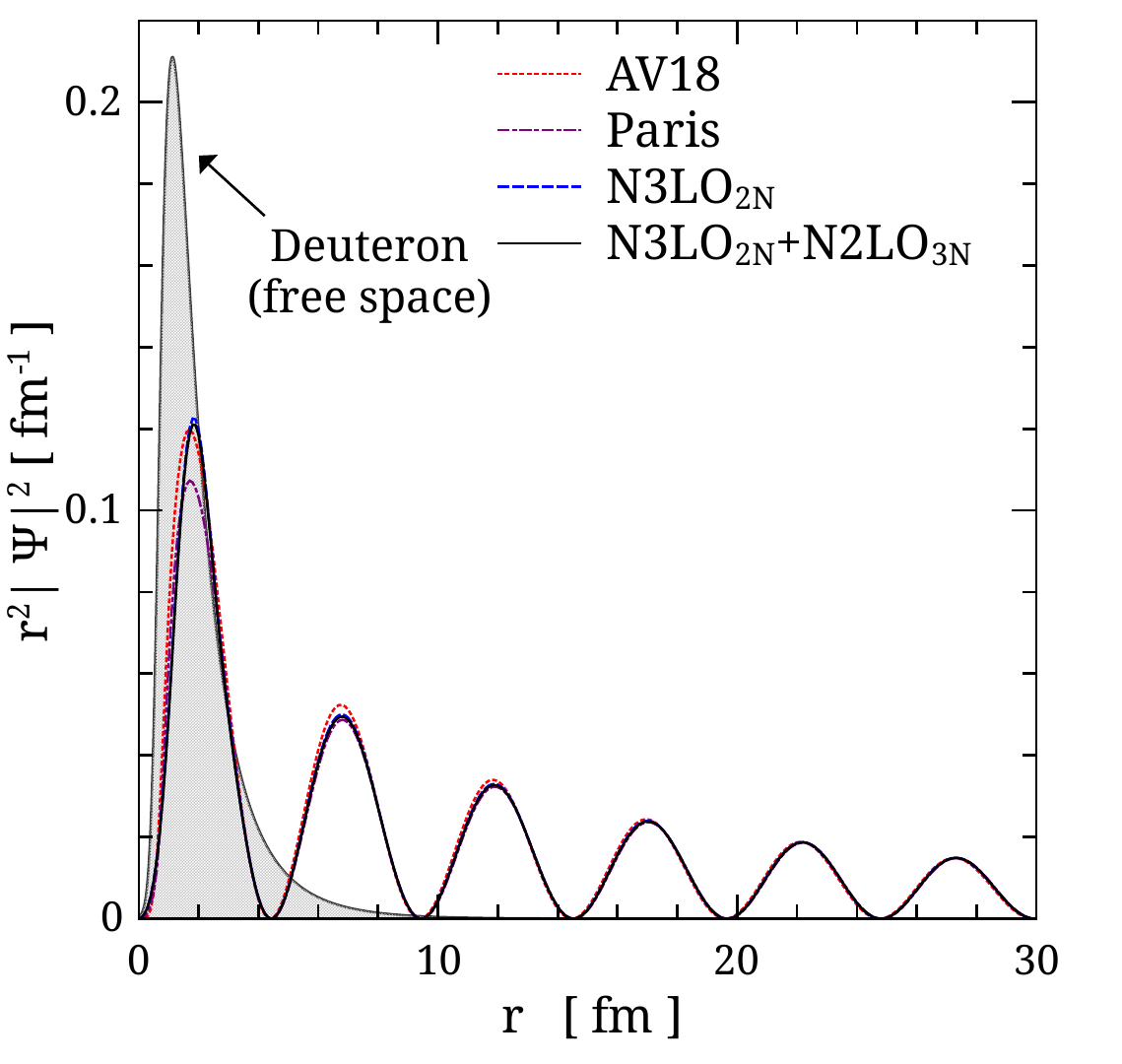}
  \caption{\label{fig_probability}
Radial probability density, $r^2|\Psi(r)|^2$, 
for di--neutron bound states in pure neutron matter
($k_F=0.6$~fm$^{-1}$) as a function of the relative distance.
Short--dashed, dash--dotted, dashed and solid curves correspond to 
AV18, Paris, N3LO$_{2N}$ and N3LO$_{2N}$+N2LO$_{3N}$ 
bare potentials, respectively. 
Gray filled curve represents the deuteron (S-- and D--waves)
probability density
in free space based on the AV18 bare potential.}
\end{figure}

In Fig. \ref{fig_probability} , we show the radial probability 
density $r^2|\Psi(r)|^2$ for \emph{in-medium} di--neutrons in 
the \SO channel at $k_F=0.6$~fm$^{-1}$. 
Short--dashed, dash--dotted, dashed and solid curves represent solutions 
corresponding to AV18, Paris, 
N3LO$_{2N}$ and N3LO$_{2N}$+N2LO$_{3N}$ solutions, respectively.  
The spatial distribution of di--neutrons is very similar for the four 
potentials considered, with small deviations only in the innermost region. 
A clear, long--range oscillatory pattern is found, as expected from 
a Pauli--blocked state in the medium. 
The nodes of the oscillations occur with a periodicity
near $5.2$~fm, and the overall spatial structure
is reminiscent of a Cooper--pair wave 
function when $\Psi(r) \sim cos(k_F r)/r$ \cite{Baldo1995,Matsuo2006}.
As we see in the following subsection,
the oscillatory behavior will give rise to large mean radii, 
a trend which has also been observed for \SO bound states in symmetric 
nuclear matter \cite{Arellano2015}. 
For reference purposes, 
we include in this plot the corresponding deuteron wave function
in free space (gray--filled curve). 
Whereas the di--neutron wave--function is modulated
by oscillatory medium effects and extends to large relative distances, 
the corresponding deuteron wave function is confined to 
within a few fm of the origin. The maximum of the
deuteron density occurs slightly below the first peak of 
the di--neutron wave function, but the overall structure of 
the two wave functions in that region is similar. 

\begin{figure}[ht!]
  \centering
  \includegraphics[width=0.8\linewidth]{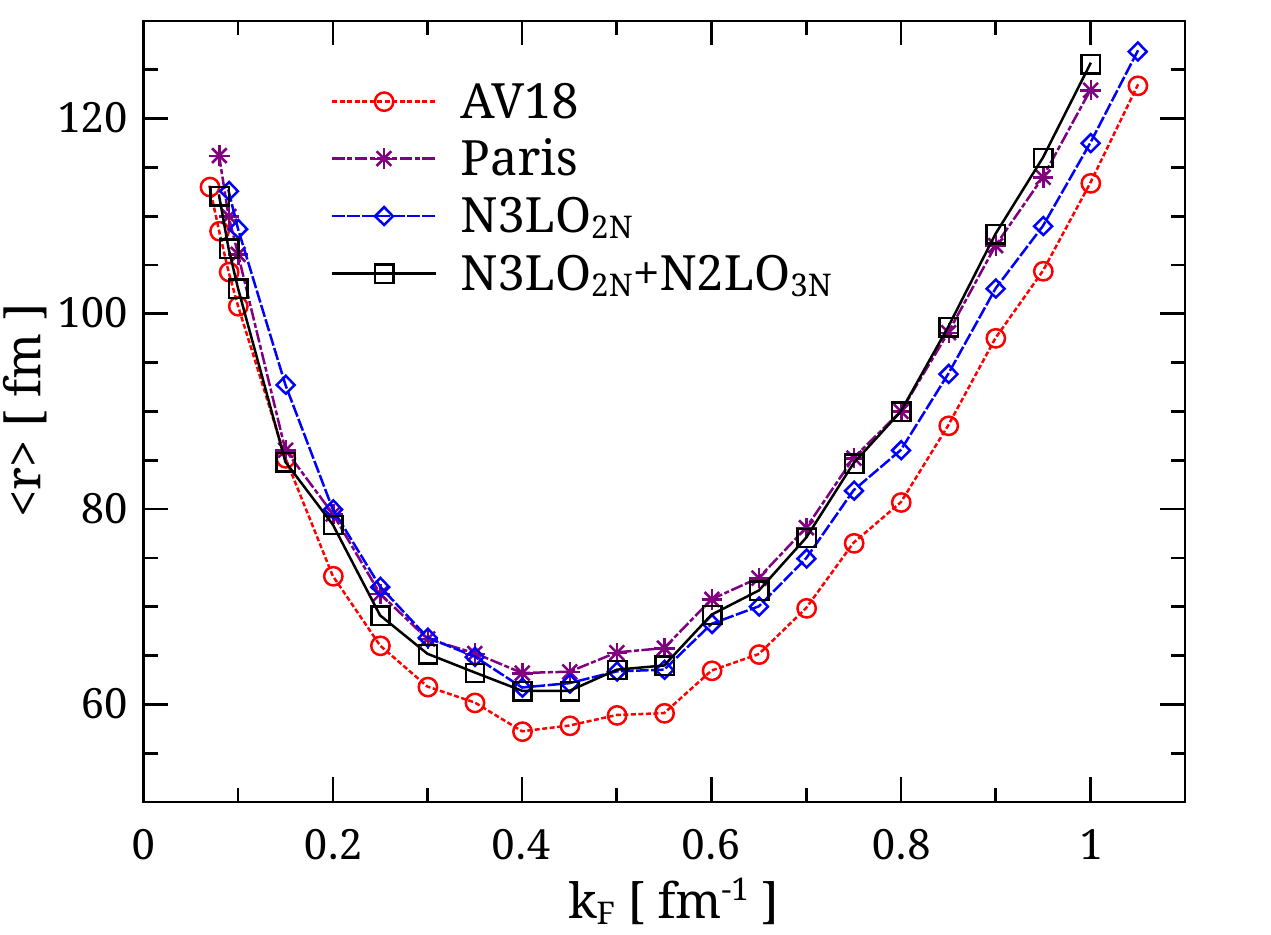}
  \caption{\label{fig_meanradii}
Mean radius $\langle r \rangle$ for di--neutrons in pure neutron 
matter as a function of $k_F$. 
Circles, asterisks, diamonds and squares 
correspond to AV18, Paris, N3LO$_{2N}$ and 
N3LO$_{2N}$+N2LO$_{3N}$ bare potentials, respectively.}
\end{figure}

Mean radii,  $\langle r \rangle$, which are
associated with the loosely bound di--neutrons, can be very large. 
In Fig. \ref{fig_meanradii} we plot $\langle r\rangle$ 
as a function of Fermi momentum for the
four interactions under consideration. 
These radii were obtained by resorting to Laplace transforms, 
as described in Ref. \cite{Arellano2015}. 
The values of these mean radii are extremely large, 
above $60$~fm in all cases. 
Their density dependence follows closely that of the binding 
energies in Fig.~\ref{fig_bnn3d}.
The more a di--neutron state gets bound, the more compact it becomes. 
In accordance, mean radii increase substantially in the regions 
where di--neutrons get diluted, and we find values of mean radii 
exceeding $110$~fm. 

In keeping with the robustness of the bound-state energies, 
the mean radii are remarkably
insensitive to the underlying nuclear interaction. The behavior of
$\langle r \rangle$ for both chiral interactions are very similar, 
pointing to a weak influence of 3NF on low--density, weakly bound 
di--neutron bound states. 
When comparing these two potentials with AV18, 
the latter yields more compact di--neutrons, by about $10$~fm.
This is to be expected in view of the slightly more attractive 
energies found in Fig.~\ref{fig_bnn}. Overall, however, the density
dependence and absolute value of $\langle r \rangle$ seems to be purely
determined by many--body correlations.

\begin{figure}[ht!]
  \centering
  \includegraphics[width=0.8\linewidth]{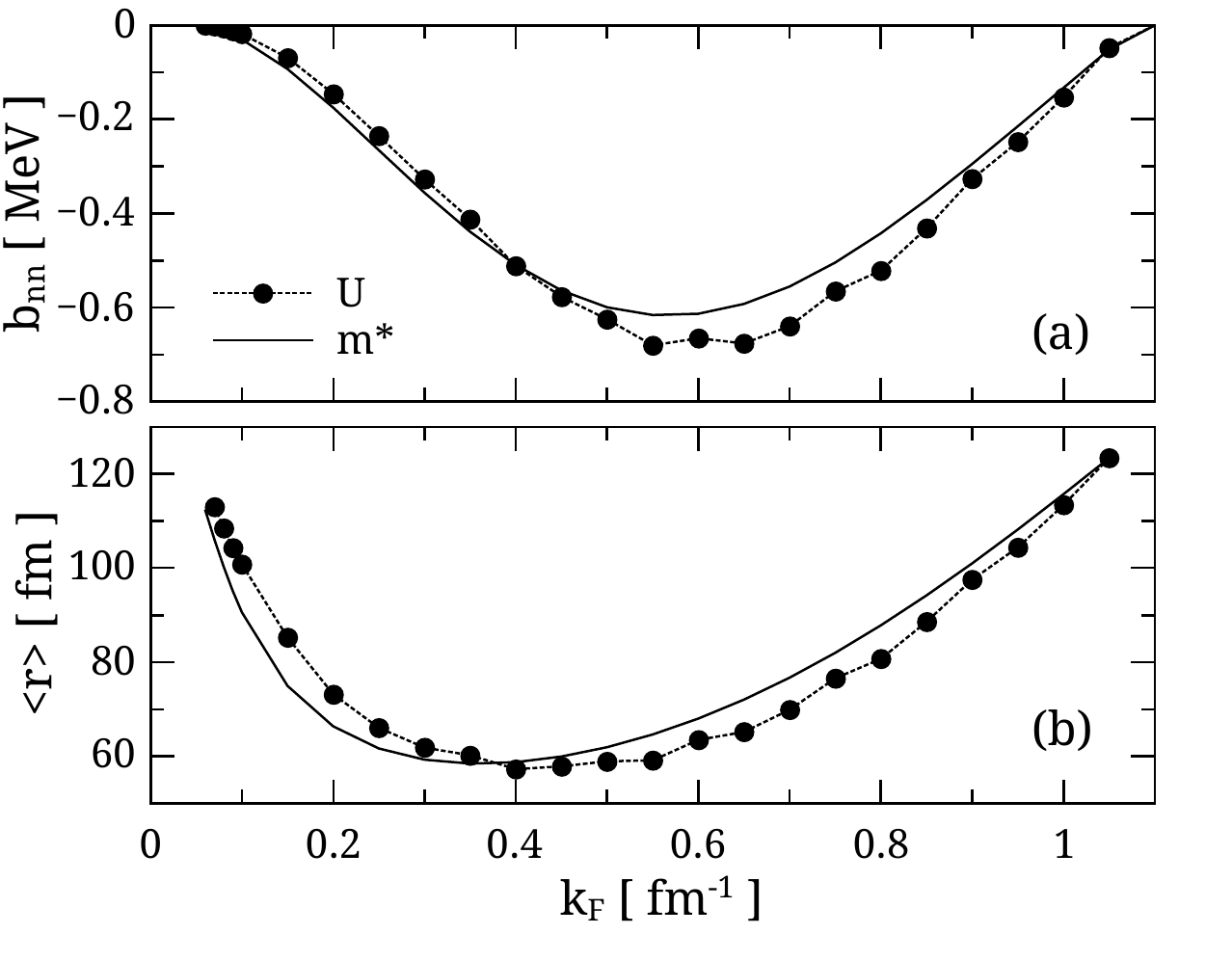}
  \caption{\label{fig_Uvsm}
(a) Di--neutron binding energy and (b) mean radii
as a function of Fermi momentum, $k_F$, for  two choices of sp spectrum.
Dashed curves with filled circles are obtained by considering the full 
spectrum, $e(k) = k^2/2m+U(k)$, whereas solid curves are obtained 
with the approximation $e(k) = k^2/2m^*$, with $m^*$ obtained 
from Eq.~(\ref{masseff}). Results are based on the AV18 bare interaction,
but similar results are obtained with other interactions.
}
\end{figure}

As discussed in Ref.~\cite{Arellano2015}, effective masses greater 
than the bare mass favor the formation of a bound state if a 
weakly attractive interaction is active. 
To explore this point in the context of pure neutron matter,
we investigate the occurrence of di--neutrons in the
\SO channel by suppressing the sp potential, while considering 
the nucleon mass equal to the effective mass at the Fermi surface, 
that is $e_k\to k^2/2m^*$, with $m^*$ obtained from Eq.~(\ref{masseff}). 
The effective mass approximation is only employed in the 
search for di--neutrons, but not in self--consistent BHF calculation. 
Indeed, if the effective mass is the main driver for the formation 
of bound states, the results should be insensitive to the removal 
of the detailed momentum dependence when the spectrum is replaced
by the constant $m^*$.
In Figs.~\ref{fig_Uvsm}(a) and  \ref{fig_Uvsm}(b) we plot di--neutron binding energies, 
$b_{nn}$ (a),
and mean radii $\langle r\rangle$, respectively, 
following the strategy described above.
Solid curves denote results based on $e_k = k^2/2m^*$, 
with $m^*$ taken from the parametrization given in Eq. (\ref{MassPar}).
We compare these results to the exact calculations (filled circles). 
We observe that the range in $k_F$ where $b_{nn}\neq 0$ is nearly 
the same in both cases, and it also coincides with the region in 
Fig. \ref{effectivemass} where $m^*/m>1$.
Moreover, the exact and the approximate calculations yield 
qualitatively and quantitatively similar binding energies and mean radii. 
Overall, this supports our interpretation connecting
bound-state formation and relatively large effective masses. 
In this respect, effective masses greater than bare masses seem 
to favor a binding mechanism for interacting neutrons.

In terms of quantitative differences, 
the effective mass approximation produces slightly
less bound di--neutron states, 
but reproduces the overall density dependence of the 
full calculations. 
In terms of radii, we find again that the shallowest bound states,
occurring at $k_F\sim 0.06$ and $1.05$~fm$^{-1}$, 
are those which yield the most spread density distributions.
We notice, however, that an effective-mass approximation
for the sp spectrum based on exact effective masses
yields quantitatively different mean radii and
binding energies (as functions of density) relative to the
exact sp solutions. The most compact di--neutron appears
at $k_F\approx 0.3$~fm$^{-1}$, slightly below the exact calculations,
although the overall radius values are very similar.
While we do not show more results here, we note that
similar results are obtained with other \emph{NN} interactions.

\subsection{Contact with Cooper pairs and di--neutron condensate}

As described in previous sections, \emph{in-medium} 
di--neutrons of total pair momentum $K$ are identified 
as poles of $g_{K}(\omega)$ at energies below the Fermi surface.
The corresponding wave functions represent confined, 
although extended, probability distributions for the bound pairs.
For vanishing total momentum, these solutions have a close 
correspondence to Cooper pairs in interacting Fermi systems,
and the subsequent emergence of superfluid states.
To sustain this remark we have solved Cooper's equation 
\cite{Cooper1956}
given by
\begin{equation}
\label{cooper}
(2e_k - \omega)\psi(k) = -\frac{2}{\pi}
\int_{k_F}^\infty k'^2 dk' v(k,k')\psi(k')\;,
\end{equation}
where $v$ represents the bare potential, and $e_k$ the sp energy.
After discretization, this equation can be reduced to a matrix 
equation to obtain its eigenenergy $\omega$
and corresponding eigenfunction $\psi(k)$ of a Cooper pair. 
The condensation of Cooper pairs in fermionic systems
can be described in BCS theory, where the energy gap $\Delta(k)$
for an uncoupled \emph{NN} state satisfies
\begin{equation}
\label{bcs}
\Delta_k = -\frac{2}{\pi}\int_0^\infty k'^2\,dk' 
v(k,k')\,\frac{\Delta_{k'}}{2E_{k'}}\;.
\end{equation}
Here $E_k=\sqrt{(e_k-\mu)^2+\Delta_k^2}$, represents a 
quasiparticle energy, with $\mu$ being the chemical potential.
The normal and anomalous density distributions corresponding to
this collective condensed many--body state are given by
\begin{align}
\label{normal}
n(k)&=\frac12 \left(1-\frac{e_k-\mu}{E_k}\right),\\
\label{anomalous}
\kappa(k)&=\frac{\Delta_k}{2E_k}\;, 
\end{align}
respectively.
The chemical potential is obtained from the condition
\begin{equation}
\label{chemical}
\rho = 2\int \frac{d^3k}{(2\pi)^3} n(k)\;,
\end{equation}
with $\rho=k_F^3/3\pi^2$ being the neutron density.
Solutions to the gap equation, involving 
Eqs.~(\ref{bcs}), (\ref{normal}) and (\ref{chemical}),
can be obtained following the method introduced by 
Baldo \emph{et al.} \cite{Baldo90}.

Equation~(\ref{bcs}) for the energy gap can be recast in terms of the
anomalous density defined by Eq.~(\ref{anomalous}),
\begin{equation}
\label{bcs2}
2\sqrt{(e_k-\mu)^2+\Delta_k^2}\, \kappa(k)=
-\frac{2}{\pi}\int_0^\infty k'^2\,dk' 
v(k,k')\,\kappa(k')\;.
\end{equation}
Note that if $|\Delta_k|\ll|e_k-\mu|$, then Eq.~(\ref{bcs2}) for the
anomalous density becomes similar to Eq.~(\ref{cooper}),
where we identify $\omega=2\mu$.
An important difference between these two equations 
is the fact that the anomalous density
is nonzero below the Fermi surface, in contrast with BHF di--neutrons
and Cooper pairs at zero temperature 
whose wave functions have fully suppressed their momentum
components below the Fermi surface.
The implication of this difference is on the range of the density
distributions.

\begin{figure}[ht!]
 \centering
 \includegraphics[width=0.8\linewidth]{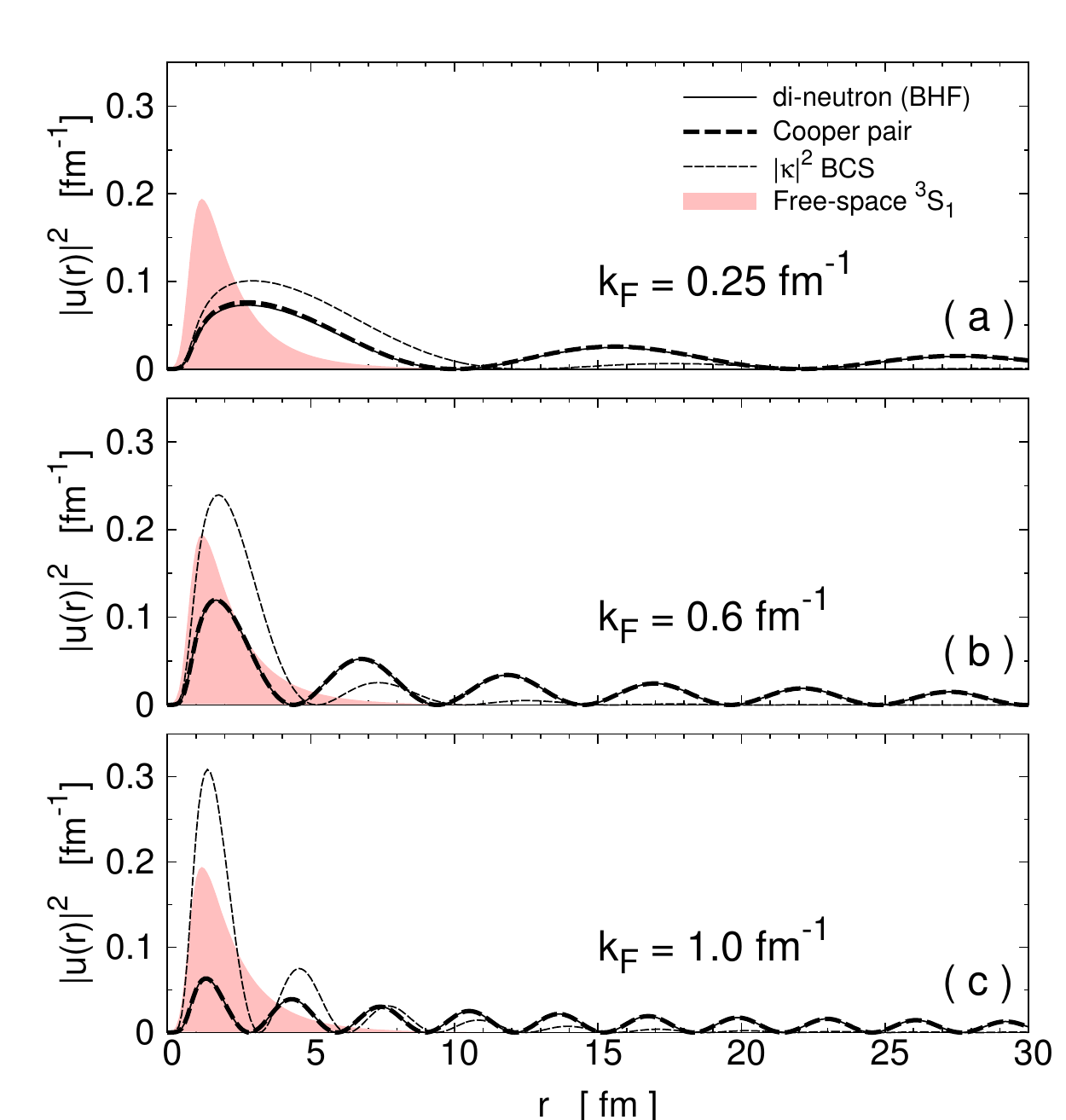}
 \caption{\label{fig_radial}
   Radial probability density for di--neutrons in $^1S_0$ channel 
   obtained from BHF equation (solid curves),
   Cooper--pair wave equation (long--dashed curves) and
   BCS anomalous density (short--dashed curves).
   Red--shaded curves in all frames represent the $^3S_1$--channel 
   deuteron radial probability density in free space.
   Panels (a)-(c) correspond to solutions for $k_F=0.25,\, 0.60,$ 
   and $1.0$~fm$^{-1}$, respectively.
 }
\end{figure}

In Fig. \ref{fig_radial} we plot the
radial probability density for di--neutrons in the $^1S_0$ channel 
obtained from the BHF equation (solid curves),
Cooper--pair wave equation (long--dashed curves) and
BCS anomalous density (short--dashed curves).
For comparison we include in all frames the $^3S_1$--channel 
deuteron radial probability density in free space,
represented with red--shaded curves.
Figures 8(a)-8(c) correspond to solutions for $k_F=0.25,\, 0.60,$ 
and $1.0$~fm$^{-1}$, respectively.
We observe that BHF di--neutron and Cooper pair density distributions are
very similar to each other, 
exhibiting long--range behavior with oscillatory
pattern with periodicity $\Delta r=\pi/k_F$.
The anomalous density distribution also shows an oscillatory pattern, 
although more confined than in the case of Cooper pairs.
This difference in size stems from the fact that Cooper pairs and 
BHF di--neutrons have a cutoff of momenta below the Fermi surface,
which is a feature that the condensate distribution does not have.
In all cases, the sizes of di--neutrons, Cooper pairs and anomalous density
appear considerably more extended than the deuteron in free space.
It is important to note at this point that Cooper pairs and
di--neutrons in BHF correspond to the same object. 
In this context, sp self--consistent potentials reported here 
account explicitly for Cooper 
pairs as described in Ref.~\cite{Arellano2015}.

\subsection{Equation of State}

In addition to its intrinsic theoretical value in the study of 
correlated nuclear systems, the energy per nucleon of pure neutron 
matter as a function of the density contains relevant information 
for the behavior of nuclear systems beyond its isospin-symmetric state. 
In the BHF approximation, the energy per nucleon is given by 
the sum of a free kinetic term and the corresponding contribution of 
the sp spectrum,
\begin{align}
\label{eq_boa2N}
\frac{B_{2N}}{A}=&
\frac{3}{10}\frac{k^2_F}{m}+\frac{1}{2}\frac{3}{k_F^3}\int_0^{k_F} 
k^2 dk \, U(k) \, .
\end{align}
This expression is valid when Hamiltonians including 2NF only are used. 
When 3NF are included in the BHF calculations, 
these enter the calculations in two ways.
First, a density--dependent two--body interaction is added 
to the bare 2NF in a standard $G$--matrix calculation. 
In addition, the total energy has to be corrected to avoid a double 
counting of the 3NF contribution \cite{Hebeler2010a,Carbone2013}. 
At the lowest order, this can be achieved
by subtracting the Hartree--Fock contribution due to 3NFs only:
\begin{align}
\frac{B_{3N}}{A}=\frac{B_{2N}}{A} &
  -\frac{1}{12}\frac{3}{k_F^3}\int_0^{k_F} k^2 dk \, 
  \Sigma^{3NF}_{HF}(k) \, .
\end{align}
We stress that the Hartree--Fock self--energy $\Sigma^{3NF}_{HF}$ 
coming from the 3N force is calculated from the effective 2N 
potential at the lowest order, in keeping with 
the procedure established in the literature \cite{Hebeler2010a}.

\begin{figure}[ht!]
  \centering
    \includegraphics[width=0.8\linewidth]{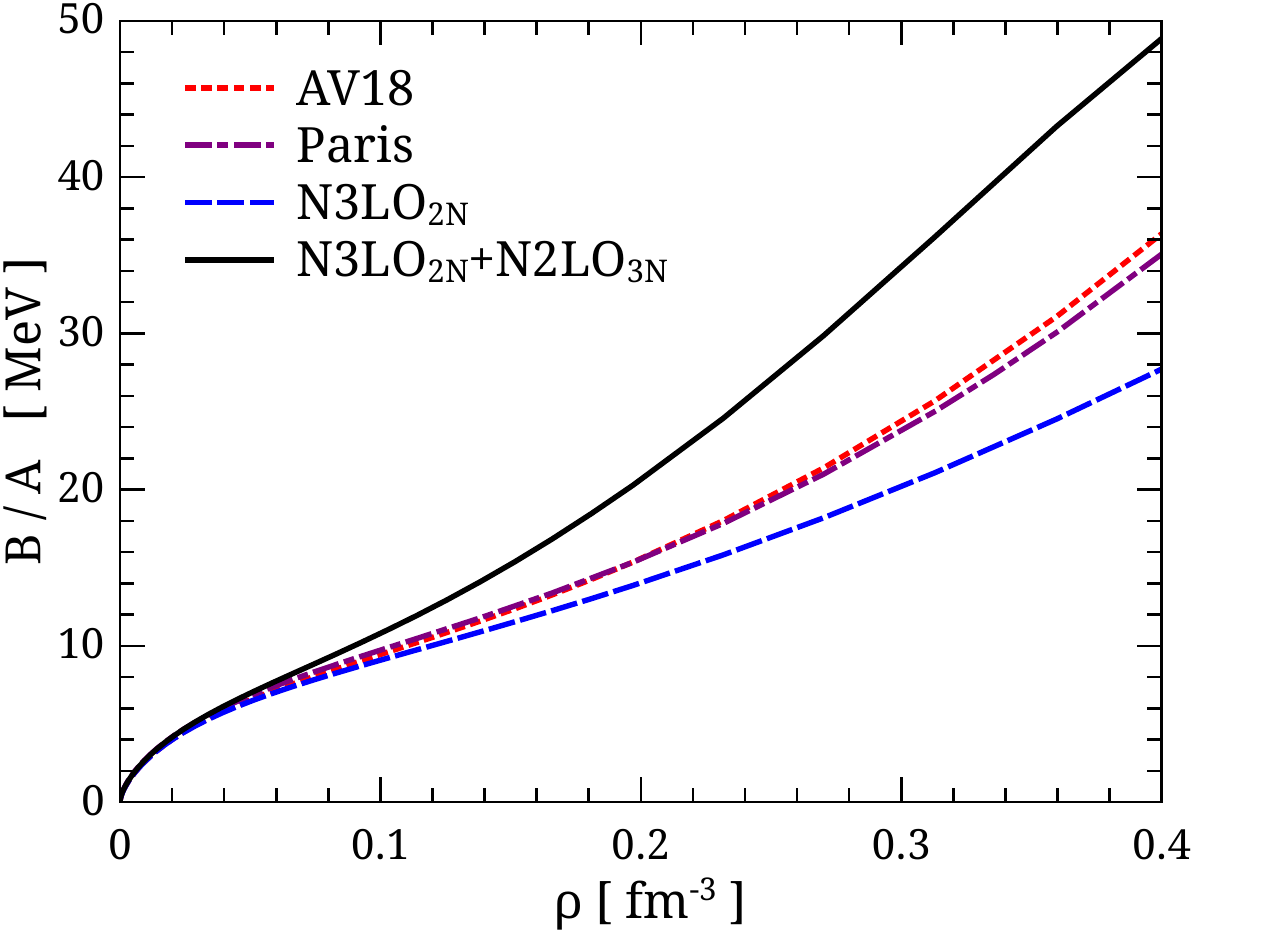}
   \caption{\label{fig_boa}
Binding energy per neutron for pure neutronic matter as a function
of density. 
Short--dashed, dash--dotted, dashed and solid curves correspond to AV18, 
Paris, N3LO$_{2N}$ and N3LO$_{2N}$+N2LO$_{3N}$ bare potentials, 
respectively. }
\end{figure}

We show the calculated energy per neutron as a function of 
density in Fig. \ref{fig_boa}. 
Again, we show results based on AV18, Paris, N3LO$_{2N}$ and 
N3LO$_{2N}$+N2LO$_{3N}$ bare interactions, 
corresponding to short--dashed, dash--dotted,
dashed and solid curves, respectively. 
For all four cases, we find the well known monotonic growth of $B/A$ 
with density. 
As expected, the case with chiral 3NFs provides the steepest increase 
in energy, with $B/A\sim 50$~MeV at $\rho=0.4$~fm$^{-3}$.
In contrast, the shallowest behavior is found for N3LO$_{2N}$, 
a very soft nuclear interaction. 
We note, in particular, that its relatively low
cutoff would normally preclude its use in densities 
well above saturation.
The other two interactions (AV18 and Paris) lay in between these
two chiral interactions. 
They provide very similar results in neutron matter, as
expected from their quantitatively similar saturation 
points and symmetry energies within the BHF approximation \cite{Li2006}.

The contribution of di--neutrons to the total energy is 
difficult to compute in this approach, but we expect it to be 
relatively small. 
First, the bound states enter the BHF calculations in
a nonlinear way, through the $G$--matrix evaluation, 
and thus their contribution is included, in a sense, 
in the energy per particle of Eq.~(\ref{eq_boa2N}). 
Second, loosely bound di--neutrons appear in a narrow 
density regime, below $\rho=0.034$~fm$^{-3}$. 
In this region, the energy of homogeneous matter is 
substantially larger and hence dominates the total contribution.
Third, an estimate of the 
contribution of di--neutrons to the total energy would require 
a calculation of the di--neutron concentration
in neutron matter \cite{Schmidt1990}, 
an issue that goes beyond our preliminary analysis.  

The results shown in Fig. \ref{fig_boa} can be summarized 
with the following functional
\begin{equation}
\label{BoAPar}
\frac{B}{A} = \frac{K}{A} + 
a \left(\frac{\rho}{\rho_0}\right)^\alpha + 
b \left(\frac{\rho}{\rho_0}\right)^\beta \, ,
\end{equation}
where $\rho_0$~$=$~$0.17$~fm$^{-3}$ is the empirical saturation density 
of symmetric nuclear matter.
We have separated explicitly the kinetic contribution, 
$K/A=3k_F^2/20m$, from the potential
terms. Coefficients of this parametrization are shown in \ref{boafit}.
\begin{table}[ht!]
\centering
\begin{tabular}{ l | c  c  c  c  }
     \hline
     & \hspace{14pt}a\hspace{14pt}  
     & \hspace{14pt}b\hspace{14pt}
     & \hspace{14pt}$\alpha$\hspace{14pt} 
     & \hspace{14pt}$\beta$ \hspace{14pt}\\ 
     & ( MeV ) & ( MeV ) & & \\
     \hline\hline
     AV18 & $5.6520$ & $-28.701$ & $1.97$ & $0.83$ \\ 
     Paris & $5.7215$ & $-28.727$ & $1.97$ & $0.87$ \\ 
     N3LO$_{2N}$ & $98.155$ & $-122.14$ & $1.13$ & $1.03$  \\ 
     N3LO$_{2N}$+N2LO$_{3N}$ & $17.767$ & $-36.815$ & $1.50$ & $0.90$  \\ 
     \hline
  \end{tabular}
  \caption{Fit coefficients for the energy per neutron, 
  $B/A$, following Eq. (\ref{BoAPar}). }
\label{boafit}
 \end{table}

\section{Summary and conclusion}

Within the BHF approach for pure neutron matter at zero temperature,
we have investigated di--neutron structures 
with emphasis placed on the low--density regime.
We have calculated self--consistent sp potentials at Fermi
momenta up to $2-3$~fm${}^{-1}$ using the continuous choice,
restricting the system to a normal (nonsuperfluid) state.
We have used AV18 and Paris internucleon potential,
in addition to chiral N3LO$_{2N}$ 2NF as well as 
N3LO$_{2N}$+N2LO$_{3N}$ 3NF.
Explicit account for di--neutron bound states is made 
in the \SO channel for the evaluation of the mass operator
during self--consistent searches of the sp potential.

The resulting sp solutions from each of the potentials considered
in this work are in fair agreement with those found 
elsewhere \cite{Li2006,Hebeler2010, Carbone2014,Baldo2012,Baldo1997}.
The major conclusion of our work is that di--neutron bound states 
appear in the BHF approach in the \SO channel independently
of the Hamiltonian that is used to model neutron interactions. 
In terms of the Fermi momentum, di--neutrons appear in the regime
 $0.06\lesssim k_F\lesssim 1.05$~fm$^{-1}$ 
 and are loosely bound, by less than $700$ keV. 
 The density dependence of their binding energy is very close for
 all interactions, which indicates a dominance of many--body
 correlations in di--neutron formation. In particular, because they
 appear at low densities, di--neutrons do not seem to be affected 
 by 3NF. More indicative, these bound states form at densities where
neutron effective masses become larger than the bare mass.
Furthermore, the size of these bound states can get
as high as $\sim\!100$~fm.
In contrast to isospin--symmetric nuclear matter, solutions for the
sp spectra do not exhibit coexisting phases \cite{Arellano2015},  
mainly due to the fact that \SD channel is fully
suppressed in pure neutron matter. 
A recent study \cite{Arellano2016} 
addressing neutronic and symmetric nuclear matter, 
considering modern realistic \emph{NN} interactions, 
confirms the robustness of theses findings.

The study of nuclear matter, 
even on its simplest nonrelativistic form,
is an exceedingly difficult problem involving various scenarios 
in density, isospin asymmetry and temperature.
On top of that, the emergence of superfluidity, superconductivity 
and clusterization multiply the physical scenarios under which
matter can evolve when confined.
The study we have presented is just a first step towards 
\emph{ab initio} clusterization studies,
going beyond traditional calculations in this context which 
are based on a variety of phenomenological approximations.
In turn, these calculations can provide guidance on clustering
in isospin--rich systems.

We can foresee a variety of extension of this work, 
both at the phenomenological and at the more theoretical level. 
Within this very same framework, we would like to study the melting 
of di--neutrons as temperature effects switch on. The extension
to isospin asymmetric systems is of relevance for neutron--star matter, 
but also in the context of nuclear structure and the isospin 
dependence of clustering correlations. The competition 
between pairing and bound states in neutron matter 
is relevant for BEC--BCS studies, and the BHF approach is perfectly 
suited to provide quantitative guidance on this subject. 
Finally, it would be interesting to use other many--body 
techniques to establish firmly the existence of di--neutron 
bound states in isospin--rich nuclear systems. 

\acknowledgments
H.F.A. acknowledges partial funding from FONDECYT under grant No 1120396. 
F.I. thanks funding from CONICYT under contract No. 221320081.
This work was supported in part by STFC through Grants ST/I005528/1, 
ST/J000051/1 and ST/L005816/1. 
Partial support comes from ``NewCompStar'', COST Action MP1304.
F.I. thanks 
the hospitality of colleagues of the University of Surrey, UK, 
where part of this work took place.

\bibliographystyle{apsrev4-1}
%\bibliography{biblio}
\end{document}